\documentclass[myepj-spec]{mySvjour}
\usepackage{graphicx}
\usepackage{hyperref}
\usepackage{color}
\usepackage{xspace}
\usepackage{natbib}   
\usepackage{verbatim}
\usepackage{amsfonts}
\usepackage{amssymb}
\usepackage{lineno}

\definecolor{darkblue1}{rgb}{0,0,.2}
\definecolor{darkblue}{rgb}{0,0,.2}
\definecolor{darkred}{rgb}{0.5,0,0}
\pagecolor{white} 
\color{black}     
\hypersetup{breaklinks=true, 
            colorlinks=true, 
            linkcolor=darkblue1, 
            menucolor=darkblue1, 
            urlcolor=darkblue1,
            citecolor=darkblue1,
            pdftitle={},
            pdfauthor={},
            pdfsubject={},
            pdfkeywords={},
            pdfproducer={}
}
\bibstyle{plain}

\newcommand{\bei}{\begin{itemize}}
\newcommand{\eei}{\end{itemize}}
\newcommand{\beq}{\begin{equation}}
\newcommand{\eeq}{\end{equation}}
\newcommand{\beqn}{\begin{eqnarray}}
\newcommand{\eeqn}{\end{eqnarray}}
\newcommand{\beqns}{\begin{eqnarray*}}
\newcommand{\eeqns}{\end{eqnarray*}}

\newcommand{\hm}{\hspace{-0.05cm}}

\newcommand{\intl}{\int\limits}

\newcommand{\mc}{\multicolumn}

\def\NIM{{ Nucl. Inst. Meth.}}
\def\ZP{{ Z. Phys.}}
\def\EPJC{{ Eur. J. Phys.}}

\def\ea{{ et al.}}
\def\sf{spectral function}
\def\sfs{spectral functions}

\def\ee{$e^+e^-$}

\def\Sew{S_{\rm EW}}

\def\bfr{branching fraction}

\def\nut{\,\nu_\tau}
\def\nueb{\,\overline{\nu_e}}

\def\to{\rightarrow}

\def\ie{{ i.e.}}

\def\rs{\raisebox{1.3ex}[-1.3ex]}

\def\@citex[#1]#2{\if@filesw\immediate\write\@auxout{\string\citation{#2}}\fi
  \@tempcnta\z@\@tempcntb\m@ne\def\@citea{}\@cite{\@for\@citeb:=#2\do
    {\@ifundefined
       {b@\@citeb}{\@citeo\@tempcntb\m@ne\@citea
        \def\@citea{,\penalty\@m\ }{\bf ?}\@warning
       {Citation `\@citeb' on page \thepage \space undefined}}%
    {\setbox\z@\hbox{\global\@tempcntc0\csname b@\@citeb\endcsname\relax}%
     \ifnum\@tempcntc=\z@ \@citeo\@tempcntb\m@ne
       \@citea\def\@citea{,\penalty\@m}
       \hbox{\csname b@\@citeb\endcsname}%
     \else
      \advance\@tempcntb\@ne
      \ifnum\@tempcntb=\@tempcntc
      \else\advance\@tempcntb\m@ne\@citeo
      \@tempcnta\@tempcntc\@tempcntb\@tempcntc\fi\fi}}\@citeo}{#1}}

\def\@citeo{\ifnum\@tempcnta>\@tempcntb\else\@citea
  \def\@citea{,\penalty\@m}%
  \ifnum\@tempcnta=\@tempcntb\the\@tempcnta\else
   {\advance\@tempcnta\@ne\ifnum\@tempcnta=\@tempcntb \else
\def\@citea{--}\fi
    \advance\@tempcnta\m@ne\the\@tempcnta\@citea\the\@tempcntb}\fi\fi}
\newenvironment{myquote}
               {\list{}{\leftmargin0cm\indent}%
                \item\relax}
               {\endlist}
\newcommand\allFontSize{\footnotesize}
\newcommand\detailsSize{\allFontSize}
\newenvironment{details}%
{\begin{myquote}\detailsSize}{\end{myquote}}

\begin{document}

\headnote{}


\title{\boldmath Update of the ALEPH non-strange spectral functions \\[0.05cm] from hadronic $\tau$ decays}

\author{M.~Davier\inst{1} \and 
        A.~H\"ocker\inst{2} \and 
        B.~Malaescu\inst{3} \and 
        C.Z.~Yuan\inst{4} \and 
        Z.~Zhang\inst{1}}
 
\institute{Laboratoire de l'Acc{\'e}l{\'e}rateur Lin{\'e}aire,
          IN2P3-CNRS et Universit\'e Paris-Sud 11, F--91405, Orsay Cedex, France \and
          CERN, CH--1211, Geneva 23, Switzerland \and
          Laboratoire de Physique Nucl\'eaire et des Hautes Energies, 
          IN2P3-CNRS et Universit\'es Pierre-et-Marie-Curie et Denis-Diderot, 
          F--75252 Paris Cedex 05, France \and
          Institute of High Energy Physics, Chinese Academy of Sciences, Beijing, China}
          
\abstract{
  An update of the ALEPH non-strange spectral functions from hadronic $\tau$ 
  decays is presented. Compared to the 2005 ALEPH publication, the main 
  improvement is related to the use of a new method to unfold the measured mass 
  spectra from detector effects. This procedure also corrects a previous problem
  in the correlations between the unfolded mass bins. 
  Results from QCD studies and for the evaluation of the hadronic vacuum polarisation 
  contribution to the anomalous muon magnetic moment are derived
  using the new spectral functions. They are found in agreement with published results
  based on the previous set of spectral functions.
}
\maketitle

\begin{flushright}
\normalsize
LAL 13-390, arXiv:1312.1501 \\
November 22, 2017 
\end{flushright}

%
%
\section{~Introduction}
\label{sec_introduction}
Because of its relatively large mass and the simplicity of its decay
mechanism, the $\tau$ lepton offers many interesting and sometimes unique
possibilities for testing the Standard Model. Among these, the production 
of hadrons from the QCD vacuum has been widely studied. 
The $\tau$ data were proved to be complementary to data from $e^+e^-$ 
annihilation, allowing one to perform detailed studies at the fundamental level 
through the determination of the spectral functions, which embody 
both the rich hadronic structure seen at low energy, and the quark behaviour 
relevant at higher energy. The spectral functions play an important 
role in the understanding of hadron dynamics at intermediate
energies and they form a basic ingredient in QCD studies and in
evaluating hadronic vacuum polarisation effects. Robust predictions of 
these effects are needed for precision tests of electroweak 
theory through the running of $\alpha$ to the $M_{\rm Z}$ 
scale, and to compute the anomalous magnetic moment of the muon. For the latter
application the spectral function of the $\pi\pi^0$ state is of 
paramount importance as it dominates the hadronic vacuum polarisation
contribution.

Following earlier determinations~\cite{aleph_vsf,aleph_asf} the ALEPH 
Collaboration released results in 2005 on the $\tau$ branching fractions and 
spectral functions based on the complete available data statistics~\cite{aleph2005}. The 
spectral function data and their covariance matrix were made public and they have been 
used in many phenomenological studies. A problem with the covariance matrix 
became apparent when fits to the spectral functions were performed~\cite{boito}. 
The statistical bin-to-bin correlations introduced by the unfolding procedure were 
not included.

After a short introduction in Section~2, we present in Sections~3 and 4 the updated 
spectral functions based on unchanged reconstructed data, but correcting 
the unfolding procedure to obtain a complete covariance matrix that properly 
includes all bin-to-bin correlations. We also utilised a new, more robust 
unfolding method~\cite{bogdan}. A fit of the $\rho$ line shape 
to the unfolded $\pi\pi^0$ data is presented in Section~5. We repeat in Section~6 the 
QCD studies performed previously~\cite{aleph2005}, and present in Section~7 
an updated evaluation of hadronic vacuum 
polarisation contributions to the muon magnetic anomaly using the $\tau$ 
spectral functions from ALEPH and other experiments. Both the QCD and hadronic
vacuum polarisation results are found in agreement with those published 
in~\cite{aleph2005}.

\section{~\boldmath Spectral functions from non-strange hadronic $\tau$ decays}
\label{sf}

The definition and determination of spectral functions are described in 
detail in Ref.~\cite{rmp}, and only a few generalities are recalled here. 
The \sf\ $v_1$ ($a_1$, $a_0$), where the subscript 
refers to the spin $J$ of the hadronic system, is defined
for a non-strange ($|\Delta S|=0$)  vector 
(axial-vector) hadronic $\tau$ decay channel ${V^-}\nut$ (${A^-}\nut$). 
The \sf\  is obtained by dividing the normalised invariant mass-squared 
distribution $(1/N_{V/A})(d N_{V/A}/d s)$ for a given hadronic mass 
$\sqrt{s}$ by the appropriate kinematic factor
\beqn
\label{eq:sf}
   v_1(s)/a_1(s) 
   &=&
           \frac{m_\tau^2}{6\,|V_{ud}|^2\,\Sew}\,
              \frac{B(\tau^-\to {V^-/A^-}\,\nut)}
                   {B(\tau^-\to e^-\,\nueb\nut)}
              \,\frac{d N_{V/A}}{N_{V/A}\,ds}\,
              \left[ \left(1-\frac{s}{m_\tau^2}\right)^{\!\!2}\,
                     \left(1+\frac{2s}{m_\tau^2}\right)
              \right]^{-1}\hspace{-0.3cm}, \\[0.2cm]
   a_0(s) 
   &=& 
           \frac{m_\tau^2}{6\,|V_{ud}|^2\,\Sew}\,
              \frac{B(\tau^-\to {\pi^-}\,\nut)}
                   {B(\tau^-\to e^-\,\nueb\nut)}
              \,\frac{d N_{A}}{N_{A}\,ds}\,
              \left(1-\frac{s}{m_\tau^2}\right)^{\!\!-2}\,
              \hspace{-0.3cm},
\label{eq:spect_fun}
\eeqn
where $\Sew$ accounts for short-distance electroweak radiative corrections. 
Since isospin symmetry is a very good approximation for the non-strange sector,
the $J=0$ contribution to the non-strange vector \sf\ is put to zero, 
while the main contributions to $a_0$ are from the pion pole,
with $d N_{A}/N_{A}\,ds = \delta (s-m_\pi^2)$.
The \sfs\  are normalised by the ratio of the vector/axial-vector 
\bfr\ $B(\tau^-\to {V^-/A^-}\nut)$ to the \bfr\ of 
the massless leptonic, \ie, electron, channel.
The CKM matrix element $|V_{ud}|=0.97418\pm 0.00019$ is taken from 
Ref.~\cite{ckmfitter-2005}.

The measurement of the $\tau$ \sfs\  defined in Eq.~(\ref{eq:sf})
requires the determination of the invariant mass-squared 
distributions, obtained from the experimental distributions after 
correcting for the effects of measurement distortion. The unfolding 
procedure used by the ALEPH collaboration, initially~\cite{aleph_vsf,aleph_asf}
and in Ref.~\cite{aleph2005},
was based on the regularised inversion of the 
simulated detector response matrix using the Singular Value Decomposition 
(SVD) technique~\cite{svd}. The regularisation function applied
minimised the average curvature of the distribution and the optimal choice 
of the regularisation strength was found by means of the Monte Carlo (MC)
simulation where the true distribution was known.

Before unfolding the mass distributions, the $\tau$ and non-$\tau$ 
backgrounds are subtracted. In the case of $\tau$ feed-through the
MC distributions normalised to the measured branching 
fractions from ALEPH~\cite{aleph2005} are used. The
contributions from strange modes classified in the same topology are
subtracted using their MC \sfs\  normalised by the measured
branching fractions.

The systematic uncertainties affecting the decay classification 
of the exclusive modes are contained in the systematic errors of 
the measured branching fractions. Additional systematic uncertainties related 
to the shape of the unfolded mass-squared distributions, 
and not its normalisation, are also included. They are dominated by the
photon and $\pi^0$ reconstruction.

\section{~Update of the  analysis using a new unfolding method}
\label{unfold}

The unfolding technique used in this reanalysis is a simplified version 
of a method developed for more complex unfolding problems~\cite{bogdan}. 
The folding probability $P_{ij}$ of an event produced in a true mass 
bin $j$ to be reconstructed in a mass bin $i$ is computed directly in MC
simulation from the transfer matrix $A_{ij}$ (the number of events produced 
in a true bin $j$ that are reconstructed in bin $i$).\footnote{The  matrix 
of folding probabilities is related to the 
transfer matrix $A_{ij}$ by $P_{ij} = A_{ij}/\sum_{k=1}^{N}{A_{kj}}$
while the matrix of unfolding probabilities is
$P'_{ij} = A_{ij}/\sum_{k=1}^{N}{A_{ik}}$.}
Conversely, the matrix of unfolding probabilities $P'_{ij}$ indicates 
the probability for an event reconstructed in a bin $i$ to originate from the 
true bin $j$, and is also computed from the transfer matrix.
$A_{ij}$ and $P'_{ij}$ depend on the assumed true spectrum
while $P_{ij}$, which describes detector and final state radiation effects, to good
approximation does not.
The method used to unfold the mass spectra is based on the idea that if 
the MC describes well enough the true spectrum in data and 
if the folding probabilities are well simulated, the matrix of unfolding 
probabilities determined in simulation can be applied to data.

If the first condition is not fulfilled, that is if the data spectrum after unfolding
differs significantly from the true MC spectrum, several steps are
iterated in which the transfer matrix is improved by re-weighting the true MC, 
keeping the folding probabilities unchanged.
Differences between data and folded (`reconstructed') MC spectra are ascribed to
differences in the unfolded (`true') spectra. At each step of the iterative 
re-weighting process, the data--MC differences of the reconstructed spectra 
are unfolded and added to the true MC spectrum. 
Such iterative procedures can result in a significant bias in the final results
if statistical fluctuations are mis-interpreted as genuine differences between 
data and MC distributions.
The method is therefore stabilised with the use of a regularisation function 
that suppresses large fluctuations in the unfolded data.
The new unfolding method is using a weaker regularisation (based on the significance 
of the data-MC differences in each bin of the spectrum) than the SVD
approach which imposes constraints on the average curvature of the spectrum~\cite{svd}.
Therefore, the new method induces less smoothing and correlations between 
mass bins. Details on the method are given in~\cite{bogdan}.

It is important to ensure that the MC simulation correctly reproduces the 
calibration and the resolution of the observed hadronic mass, which are
dominated by the $\pi^0$/photon measurement. Specific studies are 
performed using electrons from $\tau$ decays as a function of energy, 
and corrections are applied to the simulation to  match the 
properties of the data. The systematic uncertainties have been revisited following 
these studies.

\begin{figure}[t]
  \centering
  \includegraphics[width=.49\columnwidth]{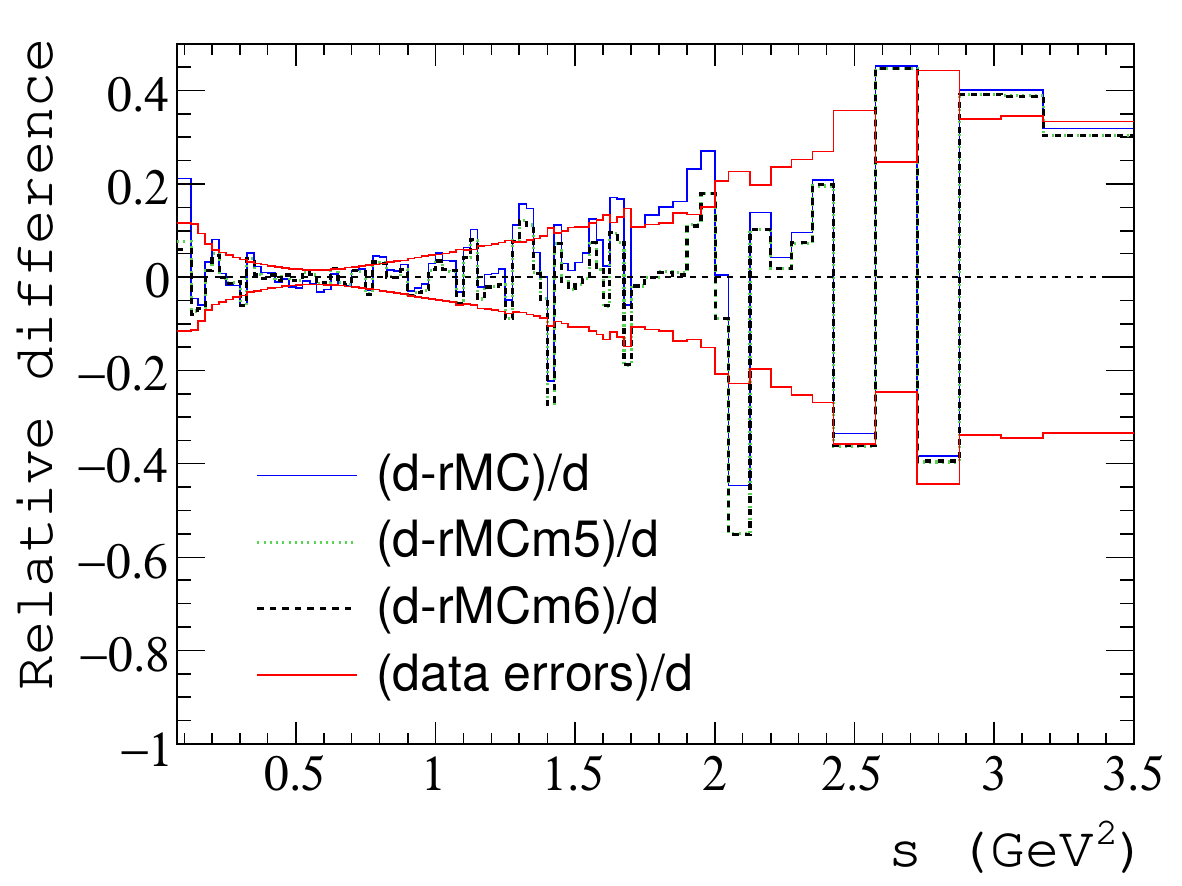}
  \includegraphics[width=.49\columnwidth]{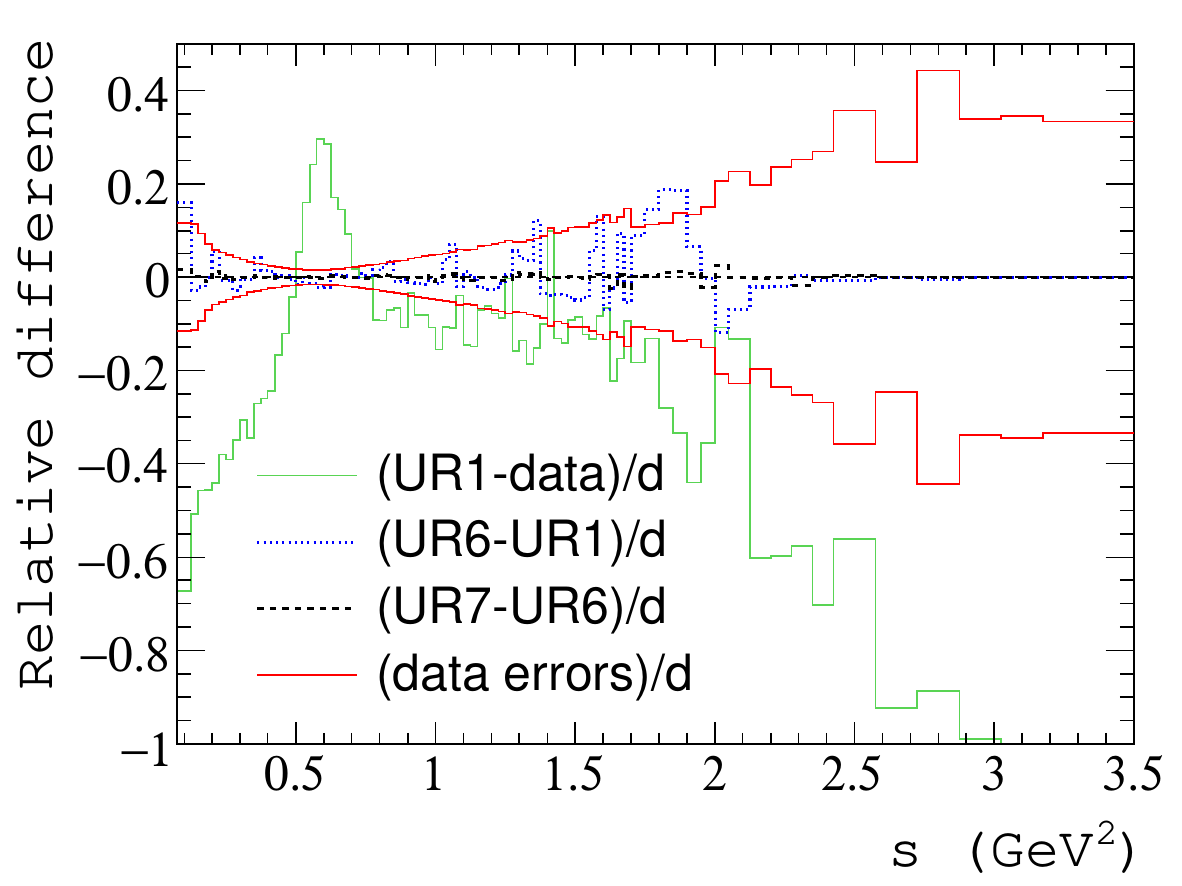}
  \vspace{0.2cm}
  \caption{\small Left: relative difference between data (d) and reconstructed MC 
                  spectrum before unfolding (rMC) and after 5 (rMCm5) and 6
                  (rMCm6) iterations for the $\pi\pi^0$ channel. Right: relative 
                  difference either between unfolded spectrum and the data or between 
                  unfolded spectra after 1 (UR1), 6 (UR6) and 7 (UR7) iterations for the same channel.}
  \label{fig:unfolding}
\end{figure}
Different numbers of bins with varying bin sizes are chosen for the unfolding 
depending on the available statistics. 
These are 83, 97, 29, 91 and 96 for the $\pi\pi^0$, $\pi2\pi^0$, $\pi 3\pi^0$, 
$3\pi$ and $3\pi\pi^0$ channels, respectively. For the vector and axial-vector 
spectra, obtained by summing the appropriate channels, a common number 
of 80 bins is adopted. The same mass-squared range up to 3.5$\;$GeV$^2$ 
is used.

The left plot in Fig.~\ref{fig:unfolding} shows the agreement between data 
and reconstructed MC for the $\pi\pi^0$ channel for different numbers of iterations. 
The agreement improves with each iteration, reaching a satisfactory level
after five iterations, the impact of further steps being very small.
The right plot in Fig.~\ref{fig:unfolding} displays the relative correction
to the measured spectrum resulting from the unfolding. Most 
of the correction is applied with the first iteration step, namely (UR1-data).
\begin{figure}[t]
  \centering
  \includegraphics[width=.55\columnwidth]{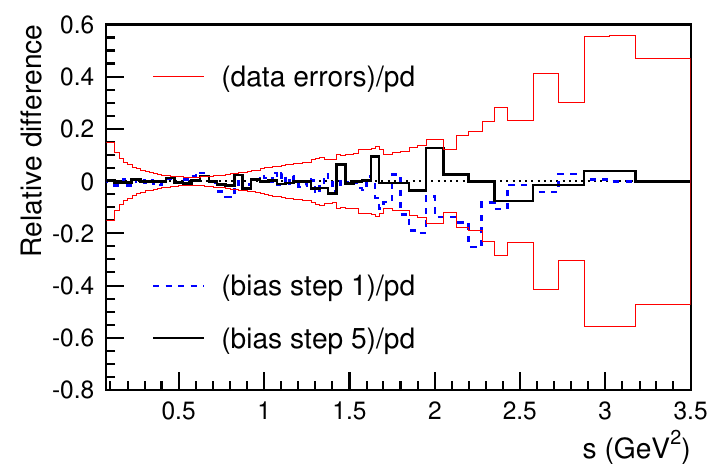}
  \vspace{0.2cm}
  \caption{\small Results from the unfolding closure test using pseudodata (pd) for the $\pi\pi^0$ channel.
                  The histograms show the relative difference (bias) between the true MC spectrum
                  and the result after unfolding of the reconstructed distribution in two cases:
                  after the first iteration (dashed) and after five iterations (solid bold). 
                  For comparison the relative statistical uncertainty on the pseudodata spectrum 
                  is indicated by the band between the two light histograms. The fluctuations in the unfolded 
                  spectrum after five iterations are of mainly statistical origin. Rebinning to broader bins reveals a
                  negligible systematic bias.}
                  \label{fig:closure}
\end{figure}

A data-driven closure test is performed to optimise the number of 
iterations and to evaluate the systematic uncertainty due to the unfolding method.
To achieve this the true MC  spectrum is reweighted using a smooth function to improve
the agreement between data and the reconstructed MC. 
The so reweighted MC spectrum is then reconstructed and provided as input to the 
unfolding process that uses the same transfer matrix as for data. The comparison 
between the reweighted true MC spectrum and the unfolded one provides a measure of the bias
introduced by the method. The results are shown in Fig.~\ref{fig:closure} for the $\pi\pi^0$ channel.
After 5 iterations the relative difference is very small and negligible 
compared to the other sources of systematic uncertainties. Inserting additional statistical fluctuations in the closure test to decorrelate the MC events used in the unfolding from those entering the response matrix does not noticeably alter the result.

\newpage
\section{~Results}
\label{results}

\begin{figure}[p]
  \centering
  \includegraphics[width=.46\columnwidth]{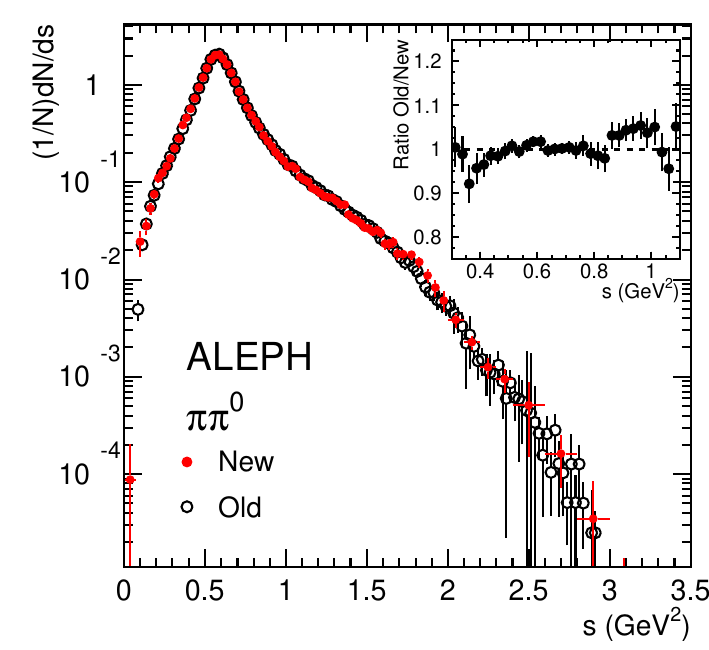}\\
  \includegraphics[width=.45\columnwidth]{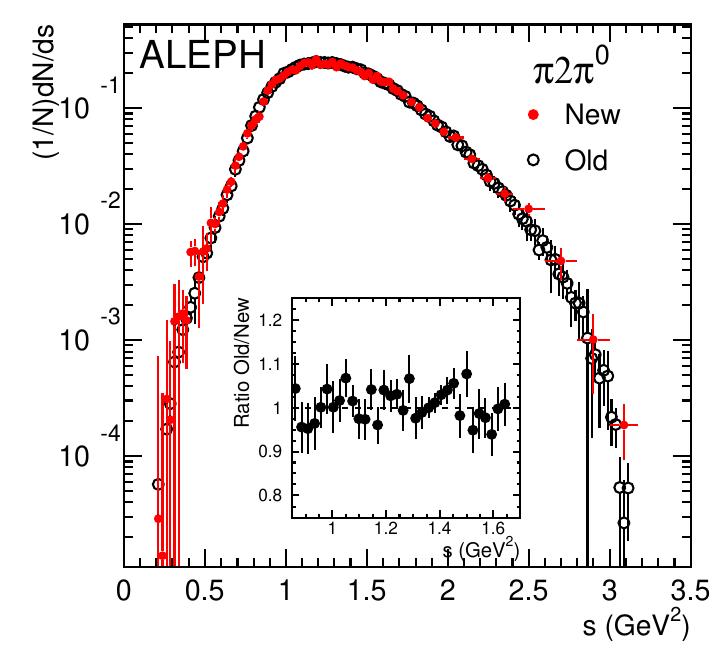}
  \includegraphics[width=.45\columnwidth]{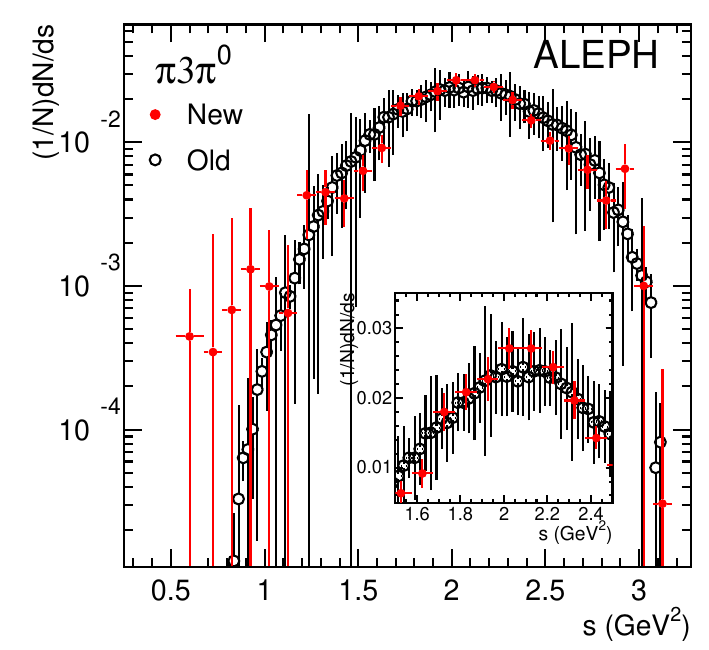}
  \includegraphics[width=.45\columnwidth]{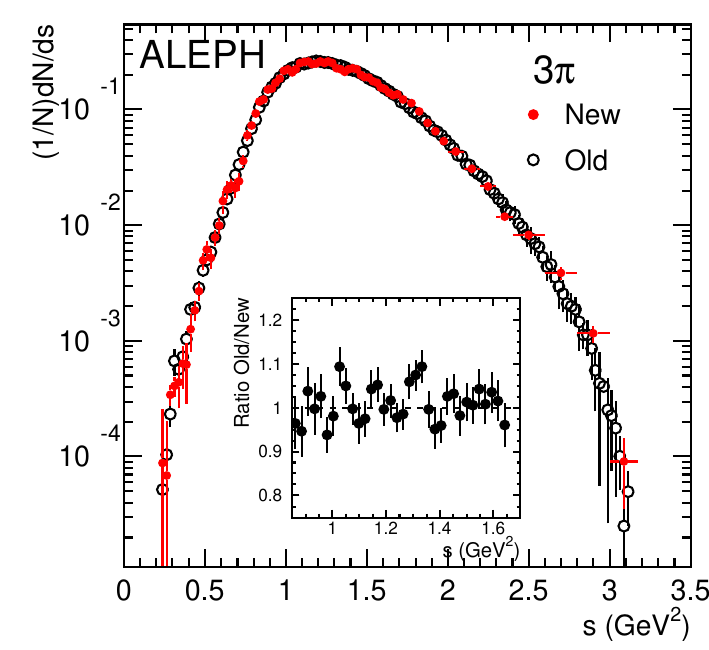}
  \includegraphics[width=.45\columnwidth]{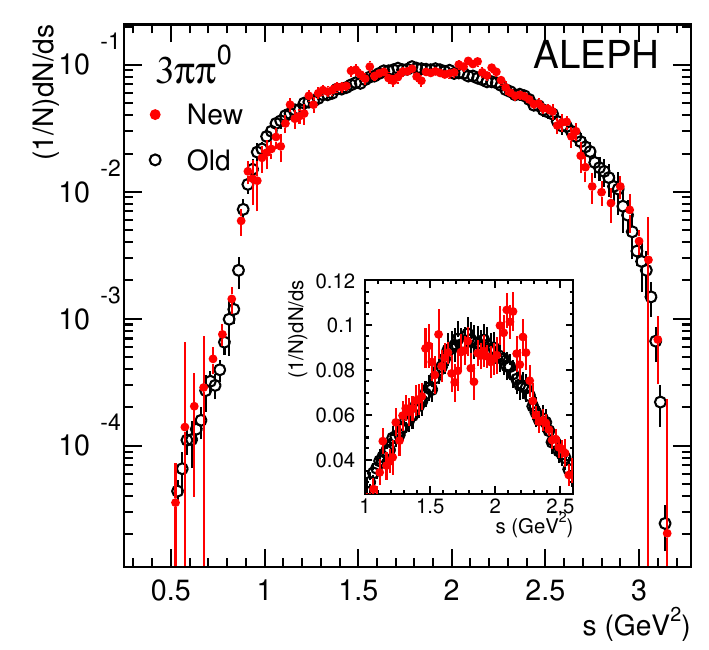}
  \vspace{0.2cm}
  \caption{\small Comparison of the new unfolded spectral functions (red full circles)
           with those obtained in Ref.~\cite{aleph2005} (black open circles, denoted ``Old"). The error 
           bars shown include statistical and all systematic uncertainties. 
           The inserts show the old-to-new ratios for better visibility, 
           where the error bars are those of the newly unfolded spectra. For the 
           $\pi 3\pi^0$ and $3\pi \pi^0$ channels the spectra are directly compared near 
           the peak since new and previous data are not given in the same energy bins.}
           \label{fig:comp_newold}
\end{figure}

A comparison of the new unfolded mass spectra with the previous 
ones~\cite{aleph2005} is shown in Fig.~\ref{fig:comp_newold}. 
Reasonable agreement is found everywhere except for  
differences at the few percent level in the $\pi\pi^0$ mode near 
threshold and in the 0.8--1.0$\;$GeV$^2$ region. 
One also observes some structures in the $3\pi\pi^0$ mode which was
not present in the previous analysis. In fact such a structure was already
there in the raw mass spectrum, but was smoothed away by our 
implementation of the SVD unfolding. An increased statistical uncertainty 
is also observed near the edges of phase space due to the reduced 
regularisation in the unfolding method employed here.

\begin{figure}[t]
  \centering
  \includegraphics[width=.46\columnwidth]{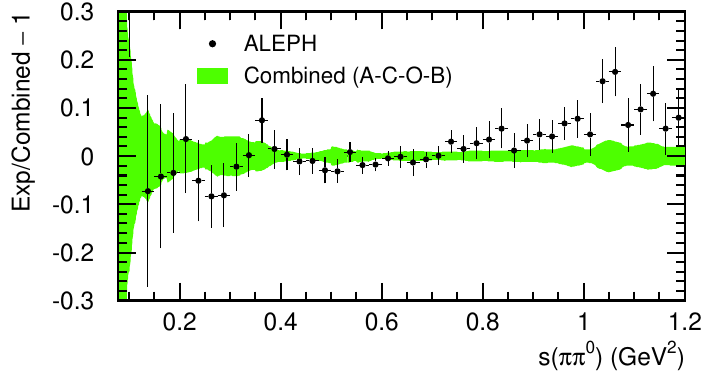}
  \includegraphics[width=.46\columnwidth]{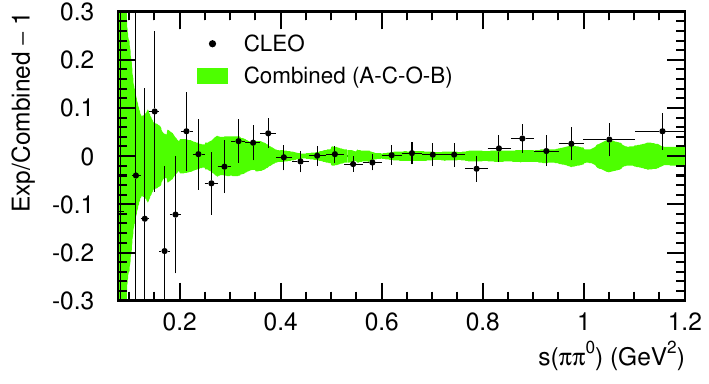}
  \includegraphics[width=.46\columnwidth]{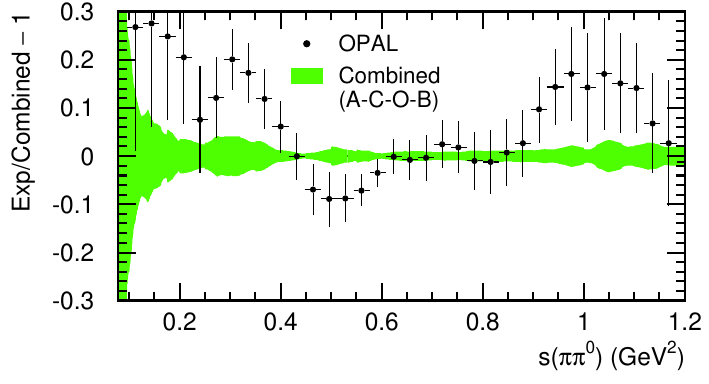}
  \includegraphics[width=.46\columnwidth]{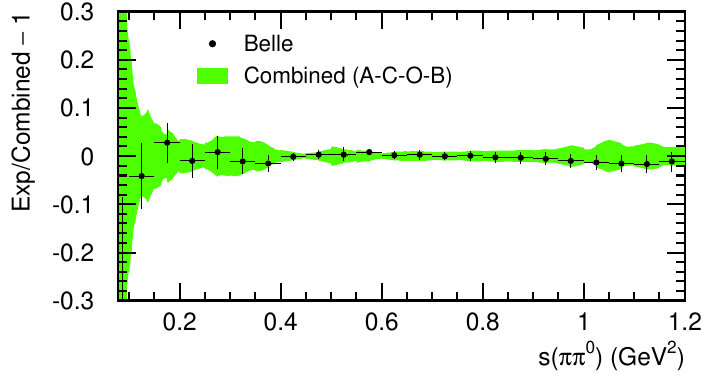}
  \vspace{0.2cm}
  \caption{\small Relative comparison between the $\tau^-\to\pi^-\pi^0\nu_\tau$
invariant mass-squared measurements from ALEPH, CLEO, OPAL and Belle 
(data points) and the new combined result (shaded band). This figure supersedes Fig.~1 of~\cite{tauee}.}
  \label{fig:comb_exp}
\end{figure}

Following the procedure defined in Ref.~\cite{tauee}, the updated ALEPH 
$\pi\pi^0$ spectral function is combined with the published results from 
CLEO~\cite{cleo-2pi}, OPAL~\cite{opal-2pi} and Belle~\cite{belle-2pi}.
The relative comparison of the individual spectral functions with the 
combination is shown in Fig.~\ref{fig:comb_exp}. It is in good agreement
with a similar comparison based on the previous ALEPH spectral functions~\cite{tauee}. 
In particular, the tension 
above 0.85$\;$GeV$^2$ between ALEPH, CLEO, and OPAL on one side and Belle on the 
other side still persists, although it is somewhat reduced with the new 
ALEPH unfolding.

A spectacular dip was found by Belle~\cite{belle-2pi} near 2.4$\;$GeV$^2$ and 
confirmed in the $e^+e^-\rightarrow \pi^+\pi^-$ cross section by
BABAR~\cite{babar-2pi}. 
As before the much lower statistics of the ALEPH data (and similarly for CLEO
and OPAL) does not permit to resolve this structure.

\begin{figure}[htbp]
  \centering
  \includegraphics[width=.49\columnwidth]{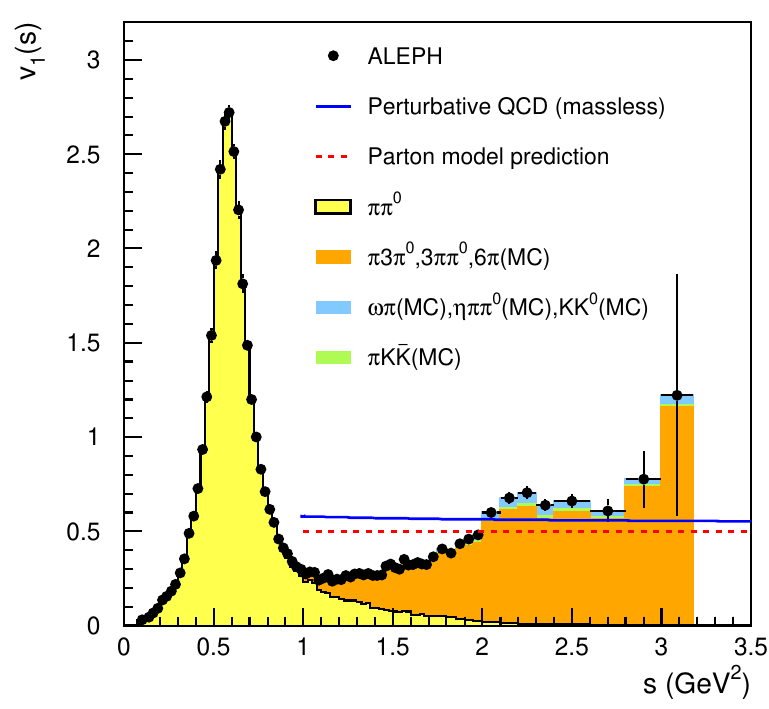}
  \includegraphics[width=.49\columnwidth]{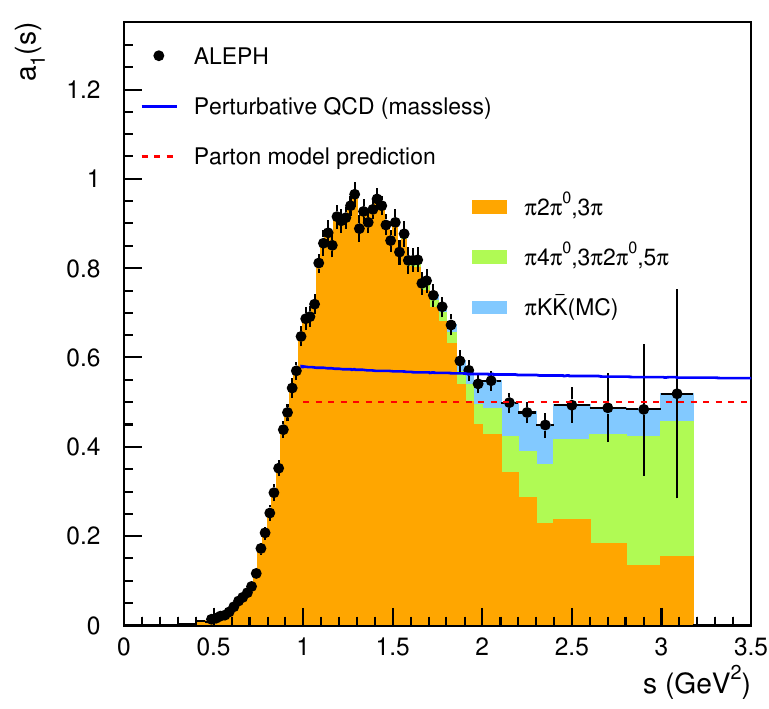}
  \includegraphics[width=.49\columnwidth]{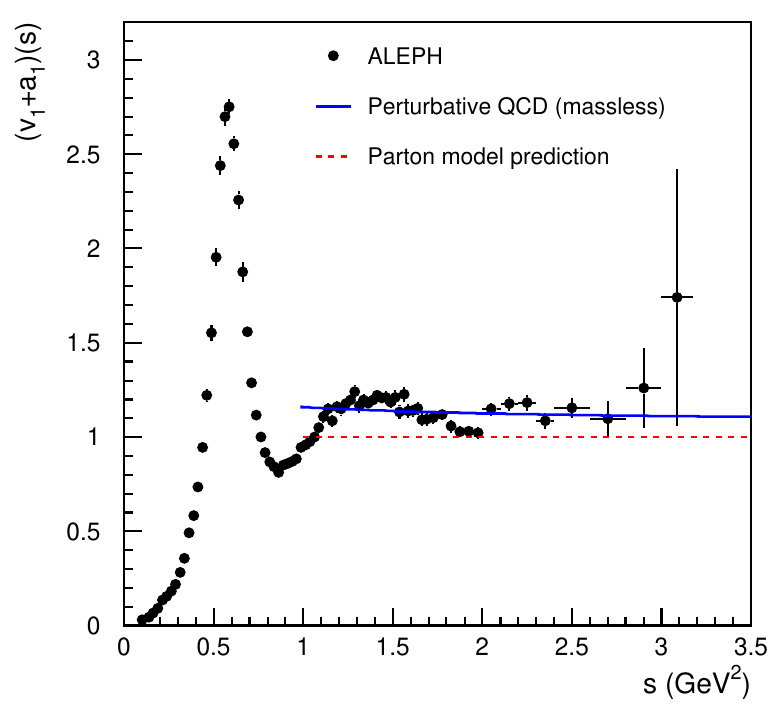}
  \includegraphics[width=.49\columnwidth]{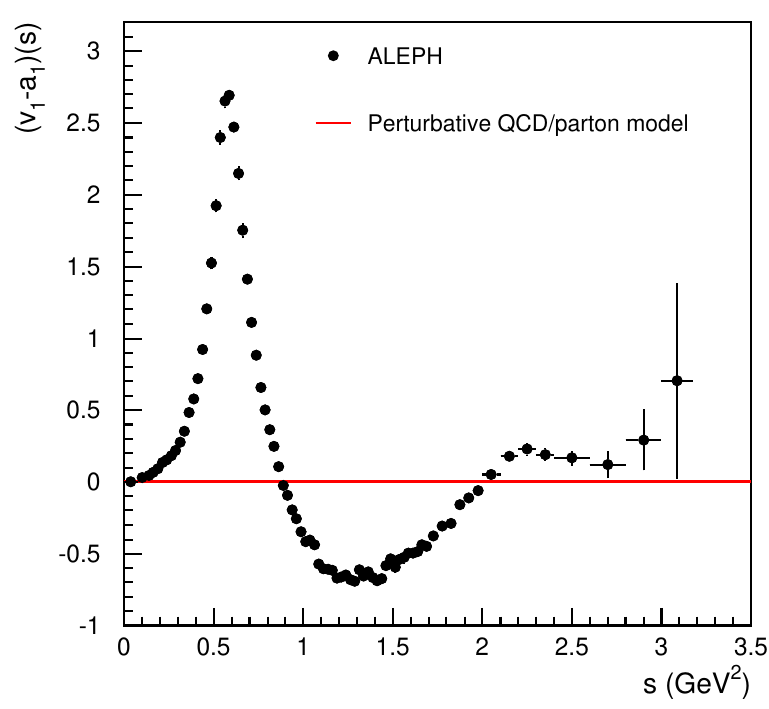}
  \vspace{0.2cm}
  \caption{\small Updated ALEPH $\tau$ spectral functions. The shaded areas indicate 
           the contributions from the exclusive $\tau$ decay channels, where the shapes 
           of the contributions labelled {\small `MC'} are taken from the MC 
           simulation. The lines show the predictions from the naive parton model and 
           from massless perturbative QCD using $\alpha_s(M_Z^2)=0.120$, respectively.
           Top left: the vector spectral function $V$. Top right: the axial-vector 
           spectral function $A$. Bottom left: the $V+A$ spectral function. Bottom right:
           the $V-A$ spectral function. This figure supersedes Figs.~62-65 of~\cite{aleph2005} and
           Fig.~2 of~\cite{alphas08}.}
  \label{fig:sf_all}
\end{figure}

\begin{figure}[htbp]
  \centering
  \includegraphics[width=.49\columnwidth]{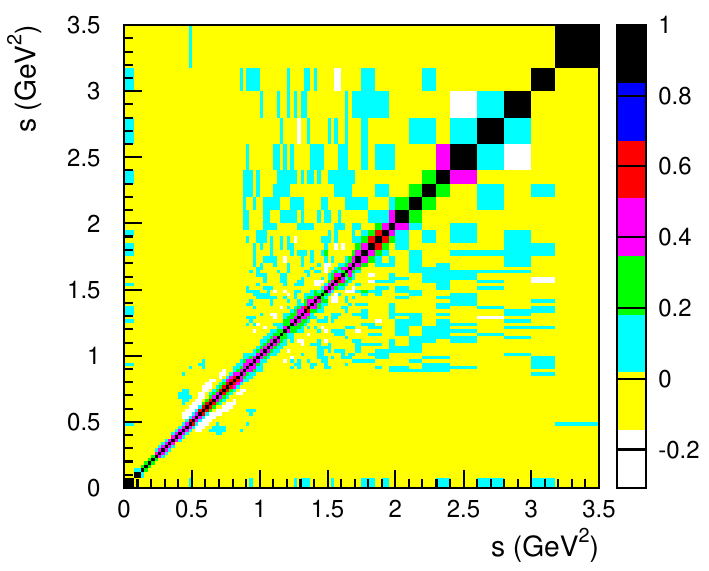}
  \includegraphics[width=.49\columnwidth]{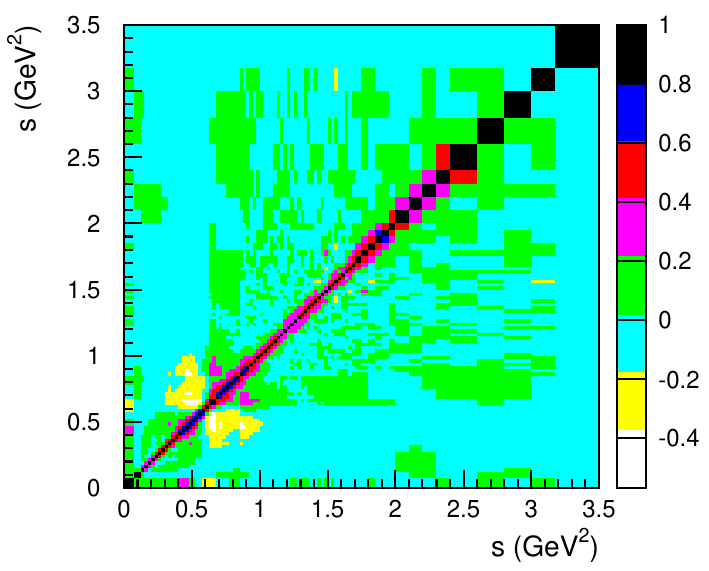}
  \vspace{0.2cm}
 \caption{Display of the correlation matrices for the vector spectral function.
          Left: statistical, with correlations induced by the unfolding. Right: 
          statistical and systematic uncertainties combined, with larger correlations and 
          anticorrelations found in the $\rho$ 
          region due to $\pi^0$ reconstruction, mass calibration and resolution effects.}
  \label{fig:covmat}
\end{figure}

The new vector ($V$), axial-vector ($A$), $V+A$ and $V-A$ spectral functions
are displayed in Fig.~\ref{fig:sf_all}. 
The correlation matrices, shown in Fig.~\ref{fig:covmat} 
for the vector part, have been carefully checked using pseudodata. 
Data for the updated spectral functions and their correlation matrices are 
publicly available~\cite{sfweb}.

\section{~\boldmath The $\rho$ line shape in the $\pi\pi^0$ channel}
\label{rhofit}

The $\pi\pi^0$ spectral function is dominated by the wide $\rho$ resonance that
can be parametrised following
Gounaris-Sakurai (GS)~\cite{gounarissak}. The statistical estimator 
minimised in the fit accounts for the correlations between different 
mass bins. 

If one assumes vector dominance, the pion form factor is given by interfering
amplitudes from the known isovector meson resonances $\rho(770)$, 
$\rho(1450)$ and $\rho(1700)$ with relative strengths 1, $\beta$, and $\gamma$.
Although one could expect from the quark model that $\beta$ and $\gamma$ are 
real and respectively negative and positive, the phase of $\beta$, $\phi_\beta$
is left free in the fits, while the much smaller parameter $\gamma$ is assumed to be 
real for lack of precise experimental information at large mass.

The parametrisation used can be found in the previous ALEPH 
paper~\cite{aleph2005}. The fitted resonance parameters given in 
Table~\ref{tab_fit_tau} are in good agreement with those obtained with 
the previous $\pi\pi^0$ spectral function, except for the $\rho(770)$ width which comes 
out larger here. The uncertainties of the fitted quantities are 
increased with the re-evaluation
of the systematic uncertainties on the mass calibration and resolution, 
and the new covariance matrix.
\begin{table}[htbp]
\begin{center}
\vspace{0.2cm}
\begin{tabular*}{\textwidth}{@{\extracolsep{\fill}}lcc} 
\hline\noalign{\smallskip}
  Parameter                   & ALEPH 2005 & This analysis  \\ 
\noalign{\smallskip}\hline\noalign{\smallskip}
$m_{\rho^\pm(770)}$ (MeV)     & $775.5 \pm 0.7$ & $775.5\pm 1.1$   \\
$\Gamma_{\rho^\pm(770)}$ (MeV)& $149.0 \pm 1.2$  & $151.4\pm 1.9$  \\ 
\noalign{\smallskip}\hline\noalign{\smallskip}
$\beta$                       & $0.120 \pm 0.008$ & $0.120\pm 0.016$ \\
$\phi_\beta$ (degrees)        & $153 \pm 7$    &  $177\pm 17$   \\
$m_{\rho^\pm(1450)}$ (MeV)    & $1328  \pm 15$   & $1404\pm 29$ \\
$\Gamma_{\rho(1450)}$(MeV)    & $468 \pm 41$  & $474\pm 84$     \\ 
\noalign{\smallskip}\hline\noalign{\smallskip}
$\gamma$                      & $0.023 \pm 0.008$ & $0.012\pm 0.022$ \\ 
$m_{\rho^\pm(1700)}$ (MeV) {\footnotesize[fixed]}   & 1713        &  1713    \\
$\Gamma_{\rho(1700)}$ (MeV) {\footnotesize[fixed]}  & 235         &  235    \\ 
\noalign{\smallskip}\hline\noalign{\smallskip}
$\chi^2$/DF                  & 119/110     &     50.4/69   \\
\noalign{\smallskip}\hline
\end{tabular*}
\end{center}
\caption[.]{Previous and new fit results of the ALEPH pion form 
            factor in the $\tau\rightarrow \pi\pi^0\nu_\tau$ channel using the Gounaris-Sakurai 
            (GS) parametrisation. The parameters $m_{\rho^\pm(1700)}$ and 
            $\Gamma_{\rho(1700)}$ are kept fixed to values obtained from 2005 fits of 
            \ee\ data extending in mass-squared up to 3.6$\;$GeV$^2$~\cite{aleph2005}.\label{tab_fit_tau}}
\end{table}

\section{~Update of the QCD analysis}
\label{qcd}

We update the QCD analysis of Ref.~\cite{aleph2005} (and references therein) 
with the new 
spectral functions. Here we follow the same notations and only briefly recall 
our method. A simultaneous f\/it of QCD predictions is performed 
including perturbative and nonperturbative components to the measured ratio 
$R_\tau$
\beq
     R_\tau = \frac{\Gamma(\tau^-\rightarrow{\rm hadrons}^-\,\nu_\tau)}
                   {\Gamma(\tau^-\rightarrow e^-\,\bar{\nu}_e\nu_\tau)}~,
\eeq
and to the spectral moments def\/ined by
\beq
\label{eq_moments}
   R_{\tau,V/A}^{kl} \;\equiv\; 
       \intl_0^{m_\tau^2} ds\,\left(1-\frac{s}{m_\tau^2}\right)^{\!\!k}
                              \left(\frac{s}{m_\tau^2}\right)^{\!\!l}
       \frac{dR_{\tau,V/A}}{ds}~,
\eeq
with $R_{\tau,V/A}^{00}=R_{\tau,V/A}$. The values for
$R_{\tau,V}=1.782 \,\pm\, 0.009$, $R_{\tau,A}=1.694 \,\pm\, 0.010$,
$R_{\tau,V+A}=3.475 \,\pm\, 0.011$, determined by the respective branching
fractions, are updated with very small changes from Ref.~\cite{alphas08}. 
Note that the $V+A$ branching fraction is obtained
as one minus the sum of leptonic and strange branching fractions.

For practical purpose, normalised moments decorrelating normalisation and 
shape information between $R_\tau$ and the spectral moments are used
\beq
\label{eq_dkl}
   D_{\tau,V/A}^{kl} \equiv
     \frac{R_{\tau,V/A}^{kl}}{R_{\tau,V/A}}.
\eeq
Their experimental values are given in Table~\ref{tab:moments} and their 
correlation matrices in Table~\ref{tab:moments_corr}. While the central values 
are in agreement with those from Ref.~\cite{aleph2005}, somewhat larger correlations 
between $R_{\tau,V/A}$ and $D_{\tau,V/A}^{10}$, and smaller correlations between higher moments, 
are observed here. 
\begin{table}[t]
\begin{center}
\begin{tabular*}{\textwidth}{@{\extracolsep{\fill}}lcccc} 
\hline\noalign{\smallskip}
Moment             & $l=0$ & $l=1$ & $l=2$ & $l=3$  \\ 
\noalign{\smallskip}\hline\noalign{\smallskip}
$D^{1l}_V$     & $0.71726$ & $0.16911$ & $0.05313$ & $0.02254$  \\
$\Delta^{\rm exp}D^{1l}_V$ & $0.00164$ & $0.00042$ & $0.00037$ & $0.00026$ \\ 
\noalign{\smallskip}\hline\noalign{\smallskip}
$D^{1l}_A$                       & $0.70940$ & $0.14885$ & $0.06586$ & $0.03191$ \\
$\Delta^{\rm exp} D^{1l}_A$ & $0.00211$ & $0.00045$ & $0.00032$ & $0.00027$  \\ 
\noalign{\smallskip}\hline\noalign{\smallskip}
$D^{1l}_{V+A}$ & $0.71343$ & $0.15924$ & $0.05934$ & $0.02710$ \\
$\Delta^{\rm exp}D^{1l}_{V+A}$ & $0.00135$ & $0.00029$ & $0.00025$ & $0.00020$ \\ 
\noalign{\smallskip}\hline
\end{tabular*}
\end{center}
\caption[.]{{Spectral moments of vector ($V$), axial-vector ($A$) and vector plus axial-vector ($V+A$) inclusive $\tau$ decays. The errors give the total experimental uncertainties including statistical and systematic effects. This table supersedes Table 23 of~\cite{aleph2005} and Table 3 of~\cite{alphas08}.\label{tab:moments}}}
\begin{center}
\vspace{0.5cm}
\begin{tabular*}{\textwidth}{@{\extracolsep{\fill}}lrrrr} 
\hline\noalign{\smallskip}
Vector             & $D^{10}_V$ & $D^{11}_V$ & $D^{12}_V$  & $D^{13}_V$\\ 
\noalign{\smallskip}\hline\noalign{\smallskip}
$R_{\tau V}$   & $-0.377$      & $0.215$        & $0.365$         & $0.389$ \\
$D^{10}_V$    & $1$                & $-0.615$       & $-0.929$       & $-0.959$  \\
$D^{11}_V$    & $-$                 & $1$                & $0.803$         & $0.597$ \\
$D^{12}_V$    & $-$                 & $-$                 & $1$                 & $0.956$ \\
\noalign{\smallskip}\hline\noalign{\smallskip}
Axial-vector             & $D^{10}_A$ & $D^{11}_A$ & $D^{12}_A$  & $D^{13}_A$\\ 
\noalign{\smallskip}\hline\noalign{\smallskip}
$R_{\tau A}$   & $-0.659$      & $0.420$        & $0.589$         & $0.594$ \\
$D^{10}_A$    & $1$                & $-0.429$       & $-0.899$       & $-0.970$  \\
$D^{11}_A$    & $-$                 & $1$                & $0.701$         & $0.414$ \\
$D^{12}_A$    & $-$                 & $-$                 & $1$                 & $0.934$ \\
\noalign{\smallskip}\hline\noalign{\smallskip}
$V+A$             & $D^{10}_{V+A}$ & $D^{11}_{V+A}$ & $D^{12}_{V+A}$  & $D^{13}_{V+A}$\\ 
\noalign{\smallskip}\hline\noalign{\smallskip}
$D^{10}_{V+A}$    & $1$                & $-0.483$       & $-0.906$       & $-0.969$  \\
$D^{11}_{V+A}$    & $-$                 & $1$                & $0.743$         & $0.508$ \\
$D^{12}_{V+A}$    & $-$                 & $-$                 & $1$                 & $0.949$ \\
\noalign{\smallskip}\hline\noalign{\smallskip}
\end{tabular*}
\end{center}
\caption[.]{{Experimental correlations between the moments $D^{kl}_{V/A/V+A}$. There 
             are no correlations between $R_{\tau, V+A}$ and the corresponding 
             moments. This table supersedes Table 24 of~\cite{aleph2005} and 
             Table 4 of~\cite{alphas08}.\label{tab:moments_corr}}}
\end{table} 

The theoretical prediction of the vector and axial-vector
ratio $R_{\tau,V/A}$ can be written as (see references and details 
in~\cite{aleph2005,rmp}):
\beq
\label{eq_delta}
   R_{\tau,V/A} \;=\;
     \frac{3}{2}|V_{ud}|^2S_{\rm EW}\left(1 + \delta^{(0)} + 
     \delta^\prime_{\rm EW} + \delta^{(2-\rm mass)}_{ud,V/A} + 
     \hm\hm\sum_{D=4,6,\dots}\hm\hm\hm\hm\delta_{ud,V/A}^{(D)}\right)~,
\eeq
with the residual non-logarithmic electroweak
correction $\delta^\prime_{\rm EW}=0.0010$, 
neglected in the following, and the dimension $D=2$ 
contribution $\delta^{(2-\rm mass)}_{ud,V/A}$ 
from quark masses which is lower than $0.1\%$ for $u,d$ quarks.
The term $\delta^{(0)}$ is the massless perturbative 
contribution, while the $\delta^{(D)}$ are the operator product expansion (OPE)
terms expressed in powers of $m_\tau^{-D}$. In Ref.~\cite{aleph2005} the 
perturbative contribution was obtained at third-order in $\alpha_s$ while 
resumming some higher-order contributions using the so-called 
contour-improved expansion in the complex energy plane. Here we take 
advantage of a more recent calculation of the fourth-order perturbative
coefficient~\cite{K4}, as we had done in a subsequent analysis~\cite{alphas08}.

The results of the fits to $R_{\tau,V/A/V+A}$ and the normalised moments to
the QCD parametrisation are given in Table~\ref{tab:fit_alphas}. Experimental
and theoretical uncertainties are separately given. Since there remains some
controversy about the proper choice of the perturbative expansion 
(fixed-order truncation, FOPT, or contour-improved method, CIPT) the final 
$\alpha_s$ results are given as the average of the two results using the $V+A$ 
spectral function. A theory uncertainty 
equal to half their difference is added. The gluon 
condensate $\langle \frac{\alpha_s}{\pi} GG\rangle$ coming from the $D=4$
contribution is separately treated, the remaining part being calculated from
the known quark masses and condensates. Table~\ref{tab:alphas_corr} provides the 
correlation matrices for the fitted parameters. Agreement is observed between the 
results in Refs.~\cite{aleph2005,alphas08} and the ones presented here.
\begin{table}[htbp]
\begin{center}
\begin{tabular*}{\textwidth}{@{\extracolsep{\fill}}lrrr} 
\hline\noalign{\smallskip}
Fitted variable & Vector $(V)$ & Axial-Vector $(A)$ & $V+A$ \\ 
\noalign{\smallskip}\hline\noalign{\smallskip}
$\alpha_s(m^2_\tau)$   & $0.346\pm 0.007\pm 0.008$      & $0.335\pm 0.008\pm 0.009$ & $0.341\pm 0.005\pm 0.006$       \\
\noalign{\smallskip}\hline\noalign{\smallskip}
$\langle \frac{\alpha_s}{\pi} GG\rangle ({\rm GeV}^4)$  & $(-0.5\pm 0.3)\cdot 10^{-2}$ & $(-3.4\pm 0.4)\cdot 10^{-2}$  & $(-2.0\pm 0.3)\cdot 10^{-2}$ \\
$\delta^{(6)}$  & $(2.8\pm 0.2)\cdot 10^{-2}$ & $(-3.7\pm 0.2)\cdot 10^{-2}$ & $(-4.6\pm 1.5)\cdot 10^{-3}$  \\
$\delta^{(8)}$  & $(-8.2\pm 0.5)\cdot 10^{-3}$    & $(10.9\pm 0.5)\cdot 10^{-3}$ & $(1.3\pm 0.3)\cdot 10^{-3}$  \\
\noalign{\smallskip}\hline\noalign{\smallskip}
$\chi^2/1 {\rm DF}$    & $0.43$                 & $3.4$                 & $1.1$    \\ 
\noalign{\smallskip}\hline\noalign{\smallskip}
$\delta^{(2)}$  & $(-3.2\pm 3.0)\cdot 10^{-4}$ & $(-5.1\pm 3.0)\cdot 10^{-4}$ & $(-4.2\pm 2.0)\cdot 10^{-4}$ \\
$\delta^{(4)}$  & $(1.0\pm 1.6)\cdot 10^{-4}$ & $(-6.3\pm 0.1)\cdot 10^{-3}$  & $(-3.1\pm 0.1)\cdot 10^{-3}$   \\
Total $\delta_{\rm NP}$  & $(2.0\pm 0.3)\cdot 10^{-2}$ & $(-3.2\pm 0.2)\cdot 10^{-2}$ & $(-6.4\pm 1.3)\cdot 10^{-3}$  \\
\noalign{\smallskip}\hline
\end{tabular*}
\end{center}
\caption[.]{{Contour-improved (CIPT) fit results of $\alpha_s(m^2_\tau)$ and the 
             OPE non-perturbative contributions for vector ($V$), axial-vector  
             ($A$) and $(V+A)$ combined fits using the corresponding ratio $R_\tau$  
             and the spectral moments as input parameters. Where two errors are  
             given the first is experimental and the second theoretical.
             The $\delta^{(2)}$ term is the pure  
             theoretical prediction with quark masses varying within their  
             prescribed range (see text). The quark condensates in the $\delta^{(4)}$  
             term are fixed to their theoretical values within uncertainties and only  
             the gluon condensate is varied as a free parameter. The total  
             non-perturbative contribution is the sum  
             $\delta_{\rm NP}=\delta^{(4)}+\delta^{(6)}+\delta^{(8)}$. 
             This table supersedes the corresponding results shown in Table 25 of~\cite{aleph2005}
             and Table 5 of~\cite{alphas08}.
             \label{tab:fit_alphas}}}
\vspace{0.2cm}
\begin{center}
\begin{tabular*}{\textwidth}{@{\extracolsep{\fill}}l|rrr|rrr|rrr} 
\hline\noalign{\smallskip}
Moment & $\langle GG\rangle_V$ & $\delta^{(6)}_V$ & $\delta^{(8)}_V$ & $\langle GG\rangle_A$ & $\delta^{(6)}_A$ & $\delta^{(8)}_A$ & $\langle GG\rangle_{V+A}$ & $\delta^{(6)}_{V+A}$ & $\delta^{(8)}_{V+A}$ \\ 
\noalign{\smallskip}\hline\noalign{\smallskip}
$\alpha_s(m^2_\tau)$ & $-0.50$ & $-0.61$ & $-0.71$ & $-0.56$ & $0.80$ & $-0.76$ & $-0.32$ & $0.55$ & $-0.64$ \\
$\langle GG\rangle_{V/A/V+A}$ & $1$ & $0.44$ & $0.71$ & $1$ & $-0.53$ & $0.78$ & $1$ & $-0.10$ & $0.54$ \\
$\delta^{(6)}_{V/A/V+A}$ & $-$ & $1$ & $0.92$ & $-$ & $1$ & $-0.92$ & $-$ & $1$ & $-0.87$ \\
$\delta^{(8)}_{V/A/V+A}$ & $-$ & $-$ & $1$ & $-$ & $-$ & $1$ & $-$ & $-$ & $1$ \\
\noalign{\smallskip}\hline
\end{tabular*}
\end{center}
\caption[.]{{Correlation matrices according to the fits presented in Table~\ref{tab:fit_alphas} 
             for vector (left table), axial-vector (middle) and $(V+A)$ (right table). As 
             the gluon condensate contributes only insignificantly to $\delta^{(4)}$, the 
             correlations to the total $\delta^{(4)}$ term are small. This table supersedes Table 26 
             of~\cite{aleph2005}. \label{tab:alphas_corr}}}
\end{table} 

The fit to the $V+A$ data using the FOPT method gives
$\alpha_s(m_\tau^2)=0.324$. Averaging with the CIPT result (see Table~\ref{tab:fit_alphas}) 
and adding to the theoretical uncertainty half the FOPT vs. CIPT difference ($\pm 0.009$)
as explained above, we find for the $V+A$ spectral function
 \beq
  \alpha_s(m_\tau^2)=0.332 \pm0.005_{\rm exp} \pm 0.011_{\rm theo}\,,
\label{final-alphas-tau}
\eeq
which after evolution to $M_Z^2$ (see Ref.~\cite{alphas08} for details) gives
\beqn
\alpha_s(M_Z^2)&=&0.1199 \pm 0.0006_{\rm exp} \pm 0.0012_{\rm theo} \pm 0.0005_{\rm evol} \\
               &=&0.1199 \pm 0.0015_{\rm tot}\,,
\label{final-alphas-Z}
\eeqn
where the third quoted uncertainty is due to the evolution.

\newpage
\section{~Update of the contribution to the muon magnetic anomaly}
\label{g-2}

The vector spectral functions are useful input to the dispersion relations
used to estimate the contribution from hadronic vacuum polarisation to the muon
magnetic anomaly. They are complementary to the spectral 
functions measured in $e^+e^-$ annihilation, but need to be corrected for
isospin-breaking (IB) effects. Here we repeat our analysis~\cite{tauee} 
using the updated ALEPH results.

\begin{figure}[t]
  \centering
  \includegraphics[width=0.6\columnwidth]{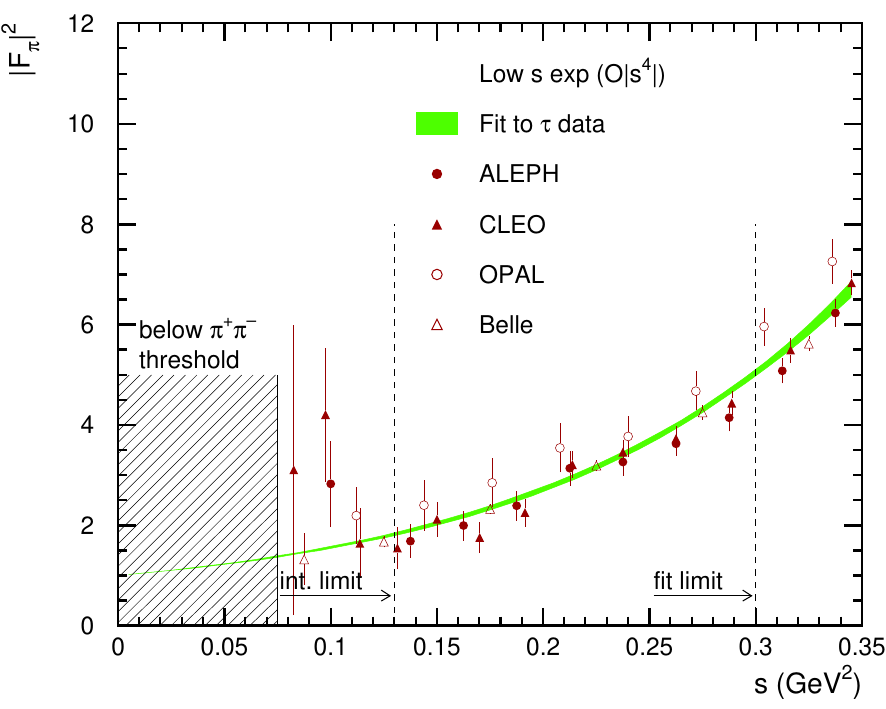}
  \vspace{0.2cm}
  \caption{Fit of the pion form factor from $4m^2_\pi$ to $0.3\;{\rm GeV}^2$ 
           using a third order expansion with the constraint $F(0)=1$ and 
           using the measured pion charge radius-squared from space-like data. 
           The result of the fit is integrated only up to $0.13\;{\rm GeV}^2$. This figure supersedes the corresponding plot in Fig.~4 of~\cite{tauee}.}
  \label{fig:fittau}
\end{figure}

In the threshold region below 0.13$\;$GeV$^2$ the data are poor, so an 
expansion constrained at $s=0$ is used~\cite{tauee}. 
The fits are shown in Fig.~\ref{fig:fittau}.
Above this value the data are directly integrated, with the results given
in Table~\ref{tab:amu_tau}. For comparison the previous ALEPH values (in 
$10^{-10}$ units) were $9.46 \pm0.33_{\rm exp}$ below 0.36$\;$GeV and
$499.2 \pm5.2_{\rm exp}$ between 0.36 and 1.8$\;$GeV, the other uncertainties 
being identical. So the new values are somewhat higher, especially near 
threshold, while the experimental uncertainties on the spectral function shape
is smaller. The latter change is not a consequence of the new unfolding, but
originates from a technical problem in the previous estimate of systematic 
uncertainties.
For the evaluation of the uncertainties
affecting the shape of the spectral functions, the normalisation of the invariant
mass spectra, given by the measured branching fractions, must be kept invariant.
This was not enforced in~\cite{aleph2005} leading to doubly assigned systematic
effects.

The results for $2\pi2\pi^0$ and $4\pi$ based on linear combinations of 
$\tau^-\to \pi^-3\pi^0\nu_\tau$ and $\tau^-\to 2\pi^-\pi^+\pi^0\nu_\tau$, 
evaluated up to $1.5\;{\rm GeV}$, are 
$14.70\pm 0.28_{\rm exp}\pm 1.01_B\pm 0.40_{\rm IB}$ and 
$7.07\pm 0.41_{\rm exp}\pm 0.48_B\pm 0.35_{\rm IB}$, respectively, 
to be compared to the previous results 
$14.89\pm 1.22_{\rm exp}\pm 1.02_B\pm 0.40_{\rm IB}$ and 
$6.31\pm 1.32_{\rm exp}\pm 0.42_B\pm 0.32_{\rm IB}$. The large difference in the
experimental uncertainties stems from the same problem in the evaluation of
systematic uncertainties mentioned above for the $\pi\pi^0$ mode.

Using the new values for the $\pi\pi^0$ mode from threshold to 1.8\,GeV combined with the data from CLEO, OPAL, and Belle and $\pi3\pi^0$ and $3\pi\pi^0$ modes below 1.5$\;$GeV, one gets a contribution from $\tau$-only input to $a^{\rm had, LO}_\mu$ 
of $537.9\pm 3.1_{{\rm exp+}B}\pm 2.0_{\rm IB}$ to be compared to the previous
value $536.4\pm 3.5_{{\rm exp+}B}\pm 2.0_{\rm IB}$. This increases the $\tau$- to 
$e^+e^-$-based~\cite{dhmz2010,hlmnt2011} difference from 1.8$\sigma$ to 2.2$\sigma$.

\section{~Conclusions}

\begin{table}[t]
\begin{center}
\begin{tabular*}{\textwidth}{@{\extracolsep{\fill}}lrr} 
\hline\noalign{\smallskip}
           &  \mc{2}{c}{$a^{\rm had, LO}_\mu[\pi\pi, \tau]$ ($10^{-10}$)} \\
\rs{Experiment}           &  \mc{1}{c}{$2m_{\pi^\pm}-0.36\;{\rm GeV}$} & \mc{1}{c}{$0.36-1.8\;{\rm GeV}$} \\
\noalign{\smallskip}\hline\noalign{\smallskip}
ALEPH &  $9.80\pm 0.40 \pm 0.05\pm 0.07$ 
      & $501.2\pm 4.5\pm 2.7\pm 1.9$ \\
CLEO  &  $9.65\pm 0.42\pm 0.17\pm 0.07$
      & $504.5\pm 5.4\pm 8.8\pm 1.9$ \\
OPAL  &  $11.31\pm 0.76\pm 0.15\pm 0.07$
      & $515.6\pm 9.9\pm 6.9\pm 1.9$ \\ 
Belle &  $9.74\pm 0.28\pm 0.15\pm 0.07$ 
      & $503.9\pm 1.9\pm 7.8\pm 1.9$ \\
\noalign{\smallskip}\hline\noalign{\smallskip}
Combined &  $9.82\pm 0.13\pm 0.04\pm 0.07$ 
         & $506.4\pm 1.9\pm 2.2\pm 1.9$ \\ 
\noalign{\smallskip}\hline
\end{tabular*}
\end{center}
  \caption[.]{{\label{tab:amu_tau}
    The isospin-breaking-corrected $a^{\rm had, LO}_\mu[\pi\pi, \tau]$ (in units of $10^{-10}$) from the 
    measured mass spectrum by ALEPH, CLEO, OPAL and Belle, and the combined 
    spectrum using the corresponding branching fraction values. The results 
    are shown separately in two different energy ranges. 
    The first errors are due 
    to the shapes of the mass spectra, which also include very small 
    contributions from the $\tau$-mass and $|V_{ud}|$ uncertainties. 
    The second errors originate from  
    $B_{\pi\pi^0}$ and $B_e$, and the third errors are due to the 
    isospin-breaking corrections, which are partially anti-correlated between
    the two energy ranges. The last row gives the evaluations using the 
    combined spectra. This table supersedes the corresponding results shown in Table 2 of~\cite{tauee}.}}
\end{table}
The ALEPH non-strange spectral functions from hadronic $\tau$
decays have been updated using a new method to unfold the measured mass 
spectra from detector effects. The new method provides a more accurate 
unfolding and corrects a problem in the correlation matrix of the 
published spectral functions~\cite{aleph2005}. The updated spectral 
functions have been used to repeat the analyses of~\cite{aleph2005}: 
a phenomenological fit to the $\pi\pi^0$ mass spectrum, a QCD analysis 
using the vector, axial-vector, and total non-strange spectral functions, and 
the computation of the hadronic contribution to the anomalous magnetic 
moment of the muon. The results obtained, although similar in most cases, 
supersede those reported in Ref.~\cite{aleph2005}.
\begin{details} 
We thank the former ALEPH Collaboration for providing the original data used
in this re-analysis.
\end{details}

%
%


\begin{thebibliography}{99}

\bibitem{aleph_vsf}  ALEPH Collaboration, \ZP\ C 76, 15 (1997).

\bibitem{aleph_asf}  ALEPH Collaboration, \EPJC\ C 4, 409 (1998).

\bibitem{aleph2005}  ALEPH Collaboration,
                     Phys. Rep.  421, 191 (2005) [hep-ex/0506072].

\bibitem{boito}      D.~Boito, private communication; D.~Boito {\it et al.},
                     Nucl. Phys. Proc. Suppl. 218, 104 (2011) [arXiv:1011.4426].

\bibitem{bogdan}     B. Malaescu, arXiv:0907.3791; \\
                     Proceedings to PHYSTAT2011 workshop, CERN-2011/006, arXiv:1106.3107.

\bibitem{rmp}        M.~Davier, A.~Hoecker and Z.~Zhang, 
                     Rev. Mod. Phys. 78, 1043 (2006) [hep-ph/0507078].

\bibitem{ckmfitter-2005}  
                     CKMfitter Group (J.~Charles \ea), 
                     Eur. Phys. J. C 41, 1 (2005) [hep-ph/0406184].

\bibitem{svd}        A.~Hoecker and V.~Kartvelishvili, 
                     \NIM\ A 372 (1996) 469 [hep-ph/9509307].

\bibitem{tauee}      M. Davier {\em et al.}, 
                     Eur. Phys. J. C 66, 127 (2010) [arXiv:0906.5443].

\bibitem{cleo-2pi}   CLEO Collaboration,
                     Phys.\ Rev.\ D 61, 112002 (2000) [hep-ex/9910046].

\bibitem{opal-2pi}   OPAL Collaboration,
                     Eur.\ Phys.\ J.\ C 7,  571 (1999) [hep-ex/9808019].

\bibitem{belle-2pi}  Belle Collaboration,
                     Phys. Rev. D 78, 072006 (2008) [arXiv:0805.3773].

\bibitem{babar-2pi}  BABAR Collaboration,
                     Phys. Rev. Lett. 103, 231801 (2009) [arXiv:0908.3589];\\
                     BABAR Collaboration,
                     Phys. Rev. D 86, 032013 (2012) [arXiv:1205.2228].

\bibitem{sfweb}      Numerical spectral function files: 
                     \url{http://aleph.web.lal.in2p3.fr/tau/specfun13.html}

\bibitem{gounarissak} G.J.~Gounaris and J.J.~Sakurai, 
                      Phys. Rev. Lett. 21, 244 (1968).

\bibitem{alphas08}   M.~Davier, S.~Descotes-Genon, A.~Hoecker, B.~Malaescu, and Z.~Zhang, \\
                     Eur. Phys. J. C 56, 305 (2008) [arXiv:0803.0979].

\bibitem{K4}         P.A.~Baikov, K.G.~Chetyrkin, and J.H.~Kuhn,
                     Phys. Rev. Lett. 101, 012002 (2008) [arXiv:0801.1821].

\bibitem{dhmz2010}   M.~Davier, A.~Hoecker, B.~Malaescu, and Z.~Zhang,
		               Eur. Phys. J. C 71, 1515 (2011) [arXiv:1010.4180].

\bibitem{hlmnt2011}  K.~Hagiwara,  R.~Liao, A.D.~Martin, D.~Nomura, and T.~Teubner, 
                     J. Phys. G 38, 085003~(2011) [arXiv:1105.3149].

\end{thebibliography}
\end{document}